\documentclass[aps,prapplied,notitlepage,superscriptaddress,showpacs,twocolumn]{revtex4-2}
\usepackage[utf8]{inputenc}
\usepackage{amsmath}
\usepackage{mathtools}
\usepackage[caption=false]{subfig}

\usepackage[ colorlinks = true,              linkcolor = blue, urlcolor  =
blue,              citecolor = red,              anchorcolor = green,
]{hyperref}

\usepackage{amssymb}
\usepackage{graphicx,color}

\usepackage{amsmath,amsthm}

\usepackage{mathrsfs}

\usepackage{bm}

\begin{document}


\title{Towards Optimal Quantum Ranging \---- \texorpdfstring{\\}{} Hypothesis Testing for an Unknown Return Signal}

\author{Lior Cohen}
\thanks{
lior.cohen3@mail.huji.ac.il}
\affiliation{Hearne Institute for Theoretical Physics and Department of Physics and Astronomy, Louisiana State University, Baton Rouge, Louisiana 70803, USA
}

\author{Mark M.~Wilde}
\affiliation{Hearne Institute for Theoretical Physics and Department of Physics and Astronomy, Louisiana State University, Baton Rouge, Louisiana 70803, USA
}
\affiliation{Center for Computation and Technology, Louisiana State University, Baton Rouge, Louisiana 70803, USA
}


\begin{abstract}
Quantum information theory sets the ultimate limits for any information-processing task. 
In rangefinding and LIDAR, the presence or absence of a target can be tested by detecting different states at the receiver.
In this paper, we use quantum hypothesis testing for an unknown coherent-state return signal in order to derive the limits of symmetric and asymmetric error probabilities of single-shot ranging experiments. 
We engineer a single measurement independent of the range, which in some cases saturates the quantum bound and for others is presumably the best measurement to approach it. 
This work bridges the gap between quantum information and quantum sensing and engineering and will contribute to devising better ranging sensors, as well as setting the path for finding practical limits for other quantum tasks.

\end{abstract}


\maketitle

\section{Introduction} One of the main goals of quantum information theory \cite{H06book,H12,Wat16,W17} is to identify the ultimate limits of information-processing tasks. One such basic task is the discrimination of two quantum states, and this problem is known as quantum hypothesis testing \cite{H69,japan1973holevo,Hel76}.

There are various ways of measuring the performance of hypothesis testing, and two prominent methods are known as symmetric and asymmetric hypothesis testing. In the former, there is an assumed prior probability distribution on the two states, and the goal is to minimize the average error probability of misidentifying the states. In the latter, no prior probability distribution is assumed, and the goal is to minimize the probability of a missed-detection error (Type~II), subject to a constraint on the false-alarm error probability (Type~I).

A sensing task, which measures an unknown physical quantity of a system, is usually accomplished by detecting a probe state that interacts with the system \cite{escher2011general}. The interaction results in a change of the probe state, and this change depends on the physical quantity of interest. Then identifying the probe state extracts the quantity's value \cite{RevModPhys.89.035002}. Thus, there is a close connection between quantum information processing and quantum sensing.

\begin{figure}[bt]
    \includegraphics[width=\columnwidth]{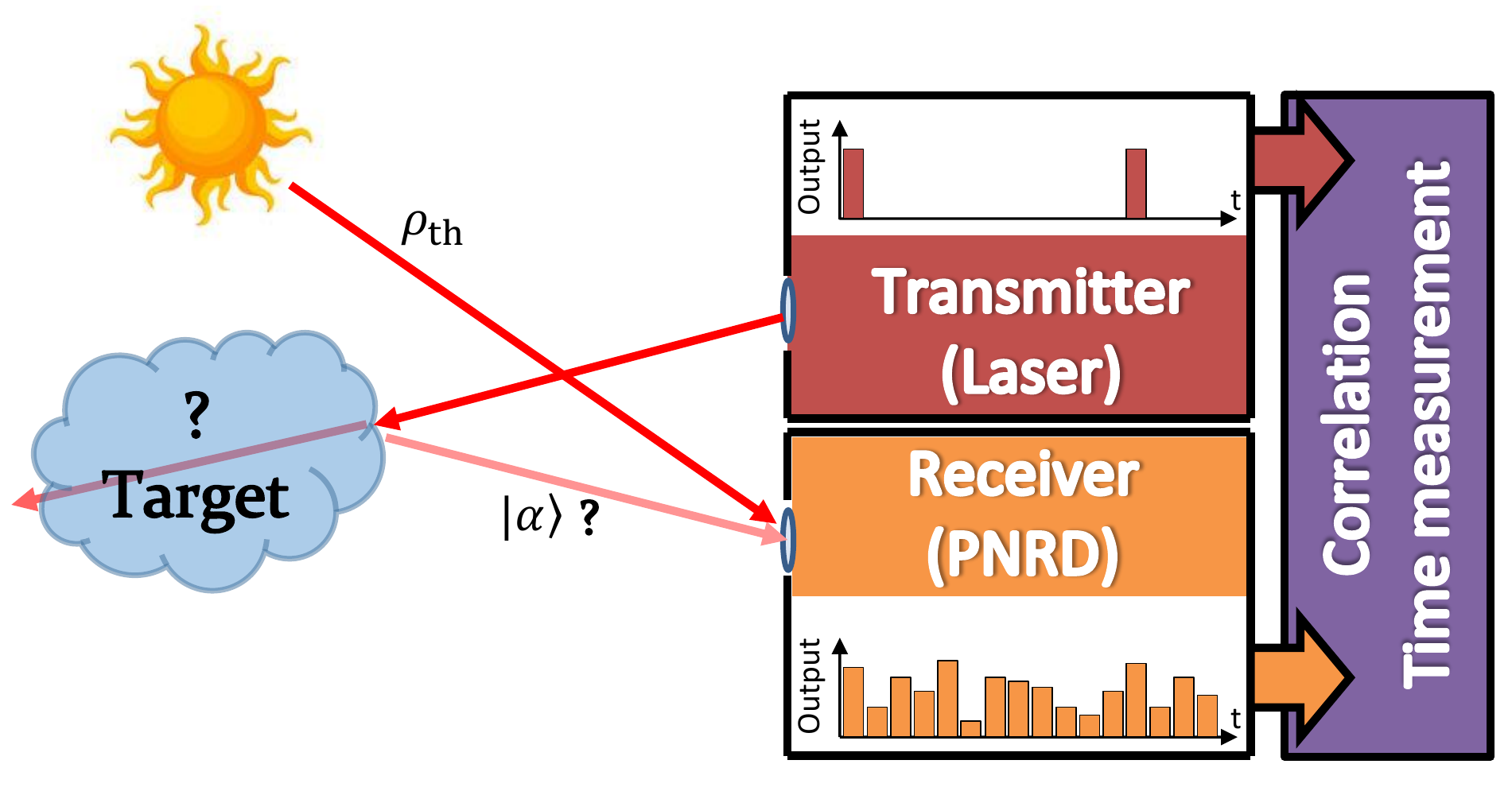}
    \caption{Concept of the presented optimal rangefinder. A short laser pulse is sent to a potential target. If the target is present, an unknown coherent state is received together with the background. If the target is absent, only the background thermal state is detected. The receiver declares a target based on the detector output then finds the range by correlating the laser’s pulse time and the target time, recorded by the detector. PNRD – photon-number-resolving detector.}
    \label{fig:system}
\end{figure}

In particular, for the well-known setting of rangefinding and LIDAR, a short laser pulse is transmitted in a direction in which a target is suspected to be present (see Figure \ref{fig:system}). If the target is present, the return state is a mixture of the original laser light and the background light, which is thermal. If the target is absent, then the return state is just thermal background light. Thus, discriminating between a thermal state and a mixture of a thermal and coherent state is equivalent to sensing the presence or absence of the target (known as target detection). Once the receiver concludes that a target is detected, the temporal information of the detector is used to find the range similarly to the regular time-of-flight ranging case. In the following we will focus our attention on state discrimination, but the results also apply to ranging by the above connection.

Target detection has been shown to be one of the tasks that can be improved by using quantum probes \cite{lloyd2008enhanced,shapiro2020quantum}. This is typically conducted, as mentioned, by differentiating between two and only two states: one representing the presence of a target and the other the absence. However, a hidden assumption is present in this commonly employed model: if the target is present, the return state is assumed to be known; in particular, its intensity and phase are assumed to be known. While the optimal detection -- the detection that saturates the quantum limit -- might be known \cite{zhuang2017entanglement} and implementable in principle, it uses all the information available about the two states. Thus, for ranging applications, in which the intensity and phase of the return state are often unknown, different limits should be found, and another detection scheme should be devised. This motivates the results reported in the present work.

In this paper, we tackle the problem of detecting an unknown return state, by proceeding in two steps. As a first step, we lift the assumption of known intensity but leave the phase fixed and known. The physical scenario corresponding to this step is one in which the range is fixed but the attenuation is unknown. As a second step, we lift the assumption of a known phase, which corresponds to the general scenario of ranging applications. More precisely, we calculate the bounds for the symmetric and asymmetric single-shot error probabilities, and we then devise a single detection scheme and measurement to minimize the error probabilities over all possible return states. To verify the theory presented here, we have also performed numerical simulations.

\section{Setup and background} In rangefinding, a short pulse of a laser is sent towards a potential target with an unknown range. The fragment of the laser light is detected by a receiver. In the absence of the laser light, a thermal state is received (corresponding to the thermal background), denoted by the state  $\rho_{\rm th}$ and defined as
\begin{equation}
    \rho_{\rm th} \coloneqq \frac{1}{1+\bar n_{\rm th}}\sum_{n=0}^\infty\left(\frac{\bar n_{\rm th}}{1+\bar n_{\rm th}}\right)^n|n\rangle\!\langle n|\,,
    \label{ThermalStt}
\end{equation}
where $\bar n_{\rm th}$ is the average photon number and $|n\rangle$ is the $n$-photon number state. We assume $\bar n_{\rm th}$ is fixed, which is usually the case. In case this assumption is not valid, one should modify the state by integrating over the intensity distribution.  Henceforth, the Fock basis is used to represent the states. (See Ref.~\cite{GK04} for background on quantum optics.)

In the presence of the target, the received state is a displaced thermal state $\rho_{\rm s}$, defined as
\begin{equation}
    \rho_{\rm s} \coloneqq \hat{D}(\alpha) \rho_{\rm th} \hat{D}(\alpha)^\dagger\,,
    \label{DisThStt}
\end{equation}
where $\hat{D}(\alpha) \coloneqq e^{\alpha\hat a^\dagger - \alpha^*\hat a}$ is the displacement operator and $\alpha=e^{i\phi}|\alpha|$ is the parameter of the unknown coherent state reflected from the target. The amplitude $|\alpha|$ is determined by the overall attenuation of the round trip to the target and back, and the parameter $\phi$ in $e^{i\phi}$ is the overall accumulated phase. In particular, both parameters depend on the range to the target. Notice that we assume here that the laser light and the background light interfere and the received light is always a single mode. The displaced thermal state $\rho_{\rm s}$ is represented in the number-state basis as follows:
\begin{multline}
     \langle n| \rho_{\rm s} |m \rangle = \\ \frac{1}{1+\bar n_{\rm th}}\sum_{k=0}^\infty \left(\frac{\bar n_{\rm th}}{1+\bar n_{\rm th}}\right)^k \langle n|\hat{D}(\alpha)|k \rangle  \langle m| \hat{D}(\alpha) |k \rangle^*\, ,
    \label{DisThStt2}
\end{multline}
and the number-state representation $\langle n |\hat{D}(\alpha)|m \rangle$ of the displacement operator $\hat{D}(\alpha)$ is as follows \cite{ivan2011operator}:
\begin{multline}
     \langle n |\hat{D}(\alpha)|m \rangle = e^{\frac{-|\alpha|^2}{2}} e^{i\phi (m-n)}\\ \times\begin{cases}  \sqrt{\frac{m!}{n!}} (-|\alpha|)^{n-m} L_{m}^{(n-m)}(|\alpha|^2) & n \ge m \\  \sqrt{\frac{n!}{m!}} |\alpha|^{m-n} L_{n}^{(m-n)}(|\alpha|^2) & m \ge n \end{cases} \,,
     \label{DispOp}
\end{multline}
where $L_{n}^{(k)}(x)$ is a generalized Laguerre polynomial. 

\section{Symmetric error setting} For a single-shot experiment, the optimal symmetric error probability is given by the Helstrom bound \cite{Hel67,H69}:
\begin{equation}
     p_{\operatorname{err}}^H = \frac{1}{2}\left(1-\frac{1}{2}\|\rho_{\rm s}-\rho_{\rm th}\|_1\right)\,,
    \label{Helstrom}
\end{equation}
where $\rho_{\rm s/th}$ are the two states under question and $\|\cdot\|_1$ is the trace norm. Related to this, the limit of a specific measurement channel $\mathcal{M}$ is given by
\begin{equation}
     p_{\operatorname{err}}^H(\mathcal{M}) = \frac{1}{2}\left(1-\frac{1}{2}\|\mathcal{M}(\rho_{\rm s}) - \mathcal{M}(\rho_{\rm th})\|_1\right)\,,
    \label{Helstrom2}
\end{equation}
where $\mathcal{M}(\omega)$ denotes the probability distribution resulting from the measurement acting on the state $\omega$ and $\left\| \cdot\right \|_1$ in this case denotes the $\ell_1$ distance of the two resulting probability distributions.

\begin{figure*}[t]
\centering
\subfloat[\label{fig:1a}]{\includegraphics[width=\columnwidth]{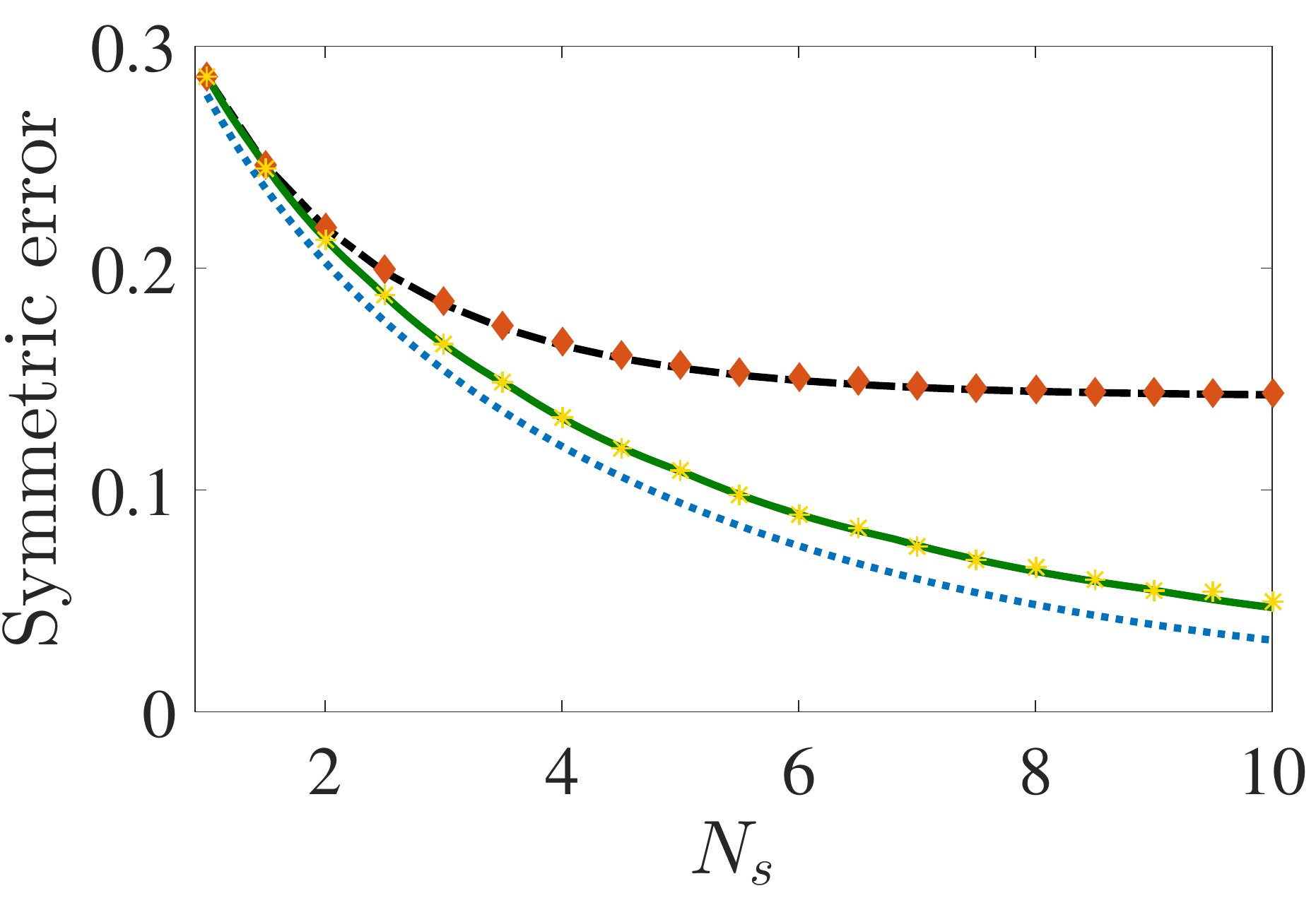}}\hfill 
\subfloat[\label{fig:1b}]{\includegraphics[width=\columnwidth]{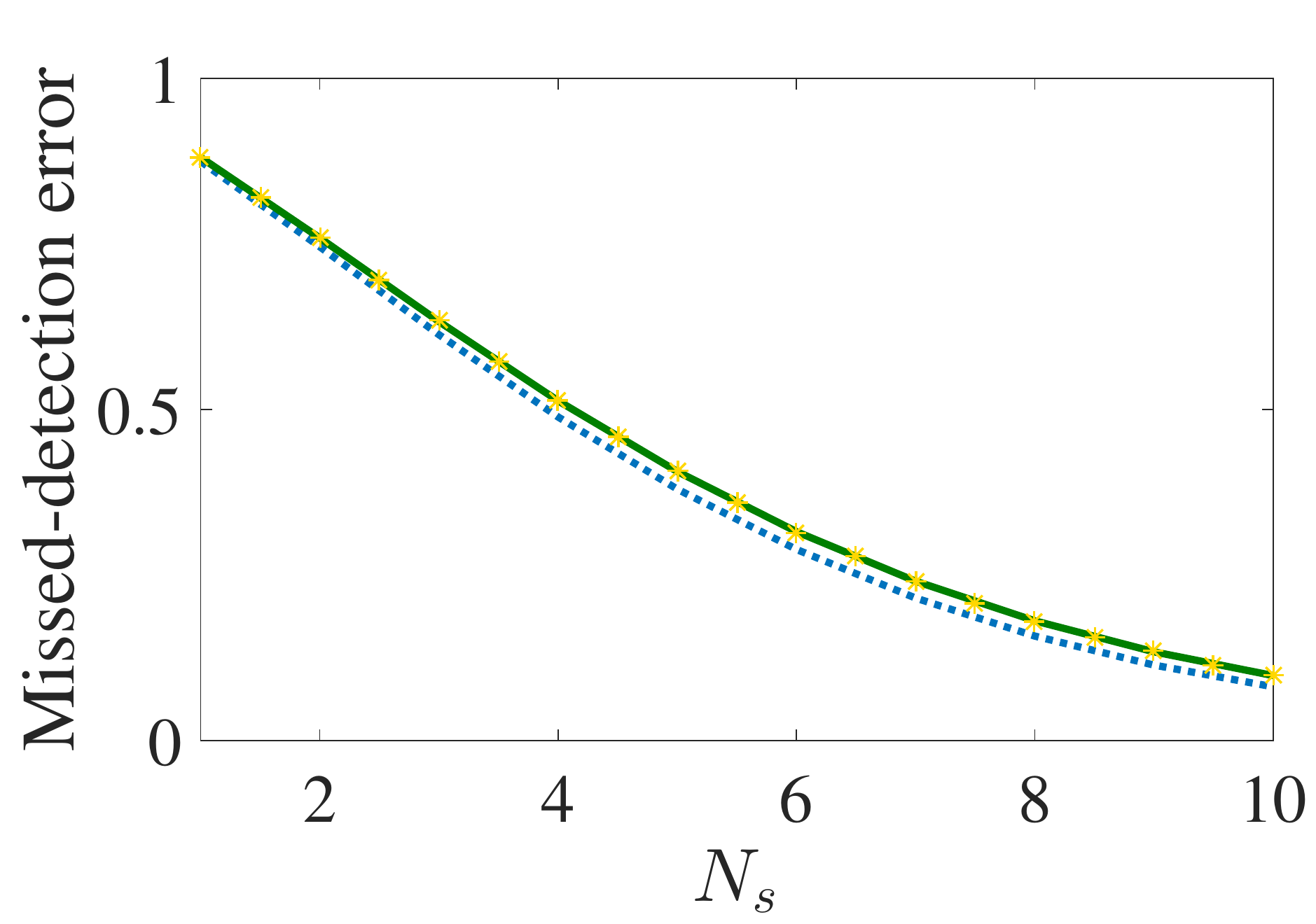}} 
\caption{Symmetric (a) and asymmetric (b) error probabilities as a function of the intensity of the return signal with a known phase. The quantum limit is shown by a blue dotted curve. The limit of photon-number detection after optimal displacement is presented with a green solid curve. The simulated data are presented with yellow stars. For the symmetric error setting, simulated data without changing the measurement are presented with orange diamonds. These data fit to the theoretical prediction (black short-dashed curve).
The average photon number of the thermal source is set to $1$. The cutoff photon number is set to 30 (see also Appendix \ref{appendix:d} and Figure~\ref{figs1} there for the cutoff dependence and Appendix \ref{appendix:c} for discussion of theory modifications for realistic detectors.
) For the asymmetric error, the false-alarm error probability threshold is set to $0.01$. See text for more details.}
\label{fig1}
\end{figure*}

The Helstrom bound as a function of the return intensity is plotted as the dotted blue curve in Figure~\ref{fig:1a}.
Notably, the intensity is related to the range and reflectivity of the target, where, for a given ranging setup (and environmental conditions), the intensity is determined solely by the target's range and reflectivity.
Reaching the Helstrom bound is accomplished by optimizing over all possible measurements. In this work, we limit the measurement setup to displacement operations followed by photon-number detection, known as the generalized Kennedy receiver \cite{kennedy1973near,takeoka2008discrimination}. This measurement scheme is experimentally realizable but does not generally achieve the Helstrom bound. It has been used for the related problem of discriminating a thermal state from a coherent state \cite{Yoshitani1970} and demonstrated error probabilities close to the Helstrom bound for that problem \cite{habif2021quantum}. More recently, this problem was addressed by employing a machine-learning approach \cite{you2020identification}.
Furthermore, photon-number detection has been demonstrated in LIDARs \cite{sher2018low}. Optimizing over displacements, the limit of this measurement setup is calculated by Eq.~\eqref{Helstrom2} and is depicted as the green solid curve in Figure~\ref{fig:1a}. We note that the displacement is optimized for $N_{\rm s}=1$ and has been kept fixed for all intensities, because we have assumed that the intensity is unknown to the receiver.

To validate the theory outlined above, we have performed numerical simulations that replicate the results of a ranging experiment. A photon number $n$ is randomly chosen, based on the probability distributions that result when the target is either absent or present. For photon number detection, the optimal decision rule is the well-known Neyman--Pearson test \cite{vT01book}:
\begin{equation}
    p_{\rm s}(n)-p_{\rm th}(n)\geq 0 ,
    \label{eq:NP-test}
\end{equation}
where $p_{\rm s}(n) \coloneqq \langle n| \hat{D}(\beta) \rho_{\rm s} \hat{D}^\dagger(\beta)|n \rangle$ is the probability of measuring $n$ photons after optimal displacement ($\beta$) with a target and $p_{\rm th}(n)\coloneqq \langle n | \hat{D}(\beta) \rho_{\rm th} \hat{D}^\dagger(\beta) |n \rangle$ after optimal displacement without a target. If the inequality in \eqref{eq:NP-test} is satisfied, the receiver concludes that the target is present, and if the inequality is violated, the receiver concludes that a target is not present. In our numerical simulations, we have repeated the detections a large number of times and counted the number of times that the receiver guesses incorrectly, leading to the average error probability presented by yellow stars in Figure~\ref{fig:1a}.  

We arrive at one immediate conclusion from the simulations: to reach the measurement limit, the photon number distribution is used. However, it is impossible to do this in a single shot if the intensity is unknown. A possible solution is to fix $p_{\rm s}(n)$ in the decision rule to be the photon number distribution for $N_{\rm s}=1$. The new theoretical prediction and the simulation results for that are also plotted in Figure~\ref{fig:1a}, as a dashed black line and orange diamonds.

\section{Asymmetric error setting} Next, we consider the setting of asymmetric error probability (see, e.g., \cite{H06book} for the general case and \cite{wilde2017gaussian} in the context of bosonic Gaussian states). In this setting, the false-alarm error probability is constrained to be below $\varepsilon \in (0,1)$,  and the missed-detection error probability is minimized subject to this constraint. 
The quantum-limited error probability is then given by
\begin{equation}
     \beta(\varepsilon) \coloneqq \min_{\hat\Lambda}\{{\rm Tr}(\hat\Lambda\rho_{\rm s}) \,|\,  0 \leq \hat\Lambda \leq I,\, {\rm Tr}(\hat\Lambda\rho_{\rm th}) \geq 1- \varepsilon \}\,,
    \label{unbalErr}
\end{equation}
where the minimization is performed over every measurement operator $\hat\Lambda$ satisfying ${\rm Tr}(\hat\Lambda\rho_{\rm th}) \geq 1- \varepsilon$. This optimization can be performed using semi-definite programming \cite{DKFRR13}, which allows for calculating $\beta(\varepsilon)$ numerically.

The ultimate limit is displayed by the dotted blue curve in Figure~\ref{fig:1b}. This limit can be modified for a specific measurement scheme $\mathcal{M}$ by replacing the states $\rho_{\rm s}$ and $ \rho_{\rm th}$ with $\mathcal{M}(\rho_{\rm s})$  and $\mathcal{M}(\rho_{\rm th})$, respectively. Taking the measurement $\mathcal{M}$ to be photon-number detection after optimal displacement, the modified limit is depicted with a green solid curve in Figure~\ref{fig:1b}.

In the case that we first perform a displacement and photon-number detection on the states involved, the optimal operator $\hat{\Lambda}$ has a simplified form, diagonal in the photon-number basis, and becomes equivalent to a stochastic decision rule. 
It takes the following form
$$
\hat\Omega = \sum_{n=0}^\infty \Omega(n) |n\rangle\!\langle n|,
$$
where $\Omega(n) \in [0,1]$ for all $n$, so that each $\Omega(n)$ corresponds to the probability of deciding that the target is absent.

\begin{figure*}[bthp]
\centering
\subfloat[\label{fig:2a}]{\includegraphics[width=\columnwidth]{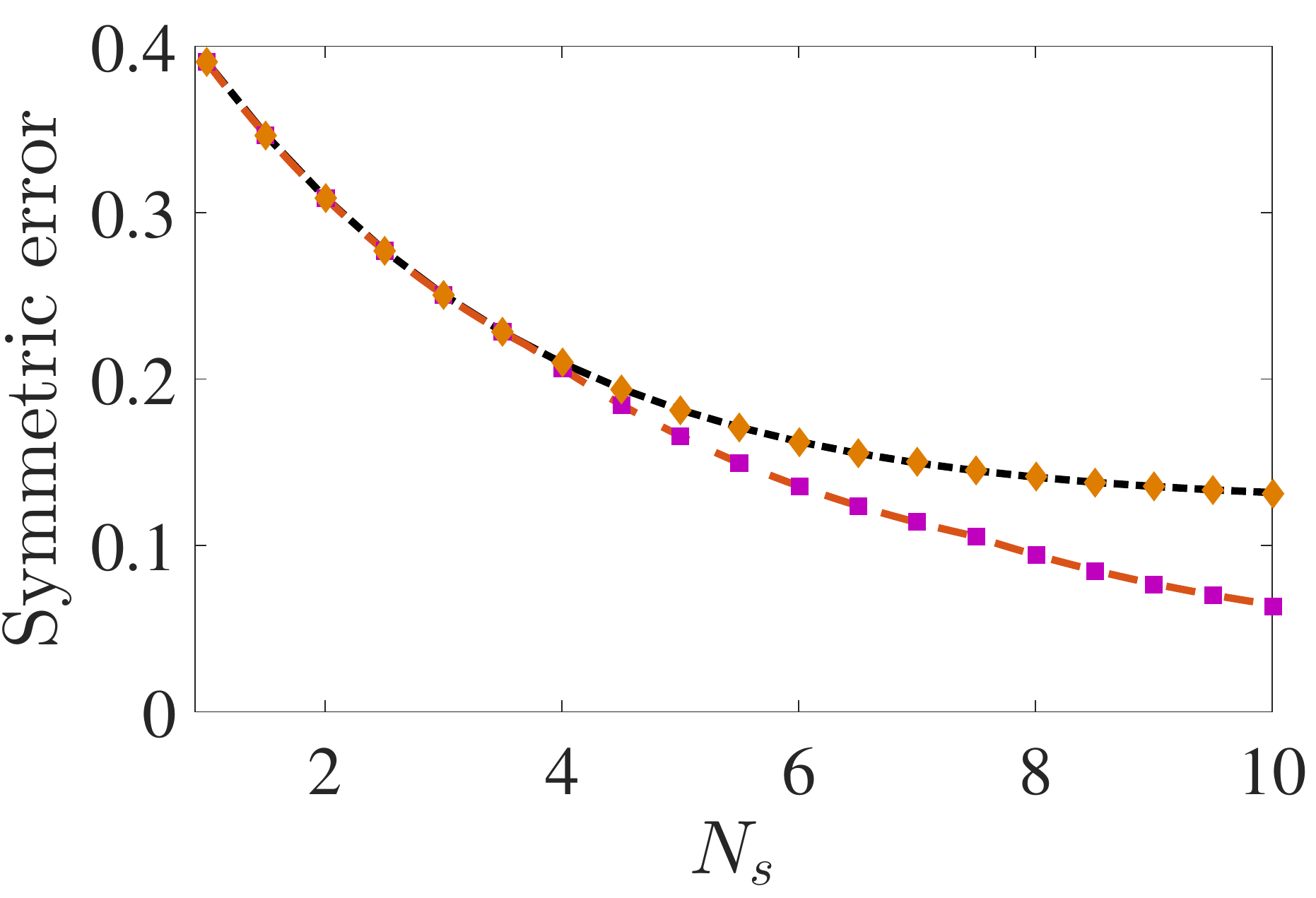}}\hfill
\subfloat[\label{fig:2b}]{\includegraphics[width=\columnwidth]{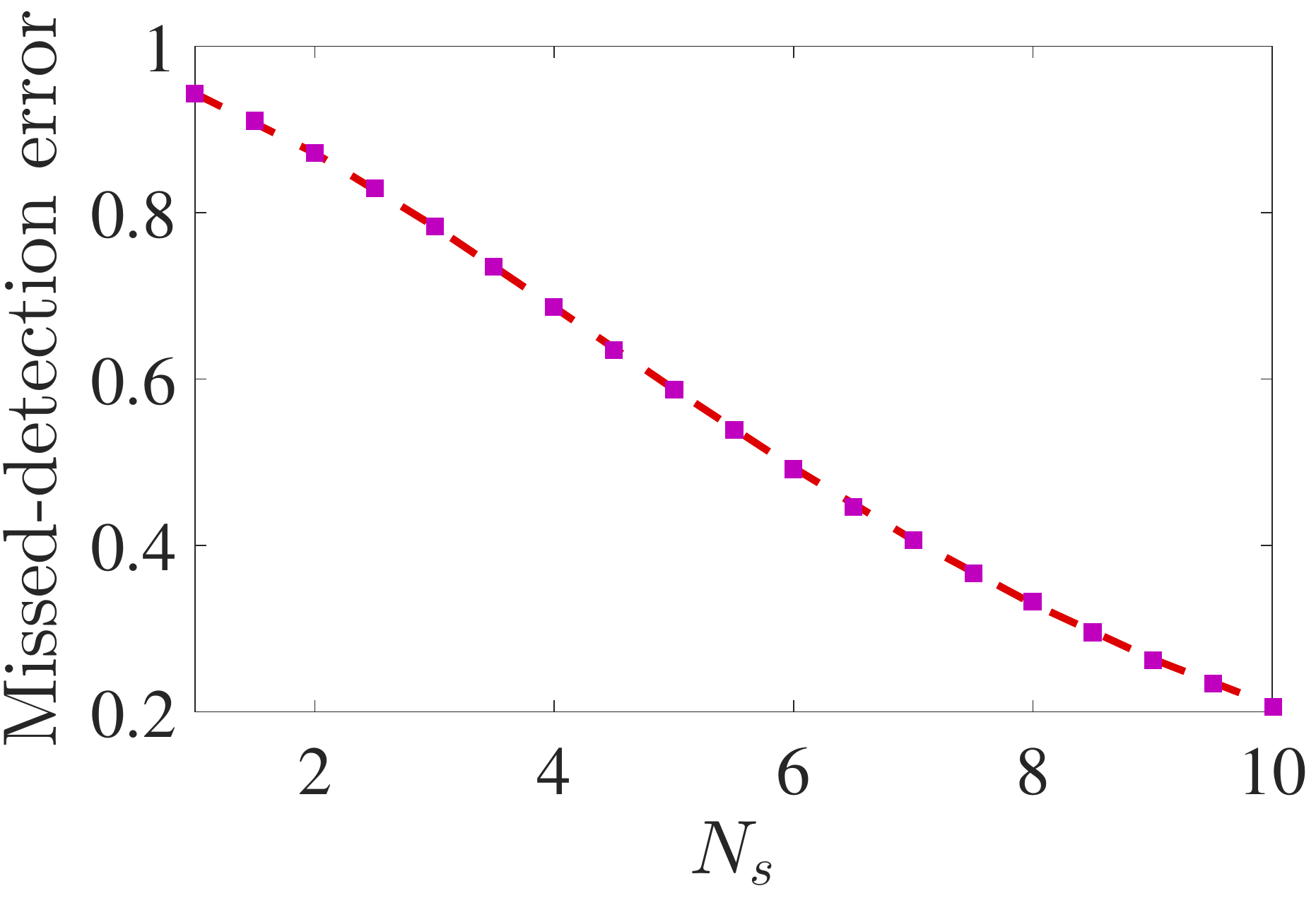}}
\caption{(a) Symmetric and (b) asymmetric error probabilities as a function of the intensity of the return signal with an unstable phase. The quantum limit is depicted as a red dashed curve. The simulated data are presented with purple boxes. For the symmetric error probability, simulated data without changing the measurement are depicted as orange diamonds. These data fit the theoretical prediction, which is depicted as a black short-dashed curve.    
The average photon number of the thermal source is set to $1$. For the asymmetric error probability, the false-alarm error probability is set to $0.01$. See text for more details.}
\label{fig2}
\end{figure*}

Using the same simulation procedure described above, we have generated simulated data for the setting involving asymmetric error probability. For a drawn photon number $n$, the receiver decides that the target is absent with probability $\Omega(n)$ and that the target is present with probability $1-\Omega(n)$. The probability $\Omega(n)$ is found by the minimization procedure given in Eq.~\eqref{unbalErr} after applying the optimal displacement and photon-number detection (see also Appendix \ref{appendix:b} and Figure~\ref{figs3} there
). The single-shot error probability is obtained by averaging the simulation results and plotted as yellow stars in Figure~\ref{fig:1b}. In this setting, unlike that for symmetric error probability, the optimal measurement is set only by the thermal state through the condition ${\rm Tr}(\hat\Lambda\rho_{\rm th}) \geq 1 - \varepsilon$. Thus, the same measurement is used for every signal intensity, and it is not affected by the unknown signal intensity.

\section{Unknown phase} Up until this point, we have considered a return signal state having a known and fixed phase. The phase was used when optimizing the displacement and when calculating the photon number distribution. In a typical single-shot ranging experiment, however, the phase is not known; thus, to calculate the average error probability for an unknown phase, one should integrate over all phases. When doing that, the results of the average errors are much worse than those in the scenario involving with known and fixed phase (see also Appendix \ref{appendix:a} and Figure~\ref{figs2} there
). This is a result of using a measurement that is phase sensitive when the phase is unknown. Moreover, in a single-shot experiment, one cannot determine if the return state has a fixed phase or an unstable phase. Thus, it can be assumed to be the latter. In this case, the signal state is
\begin{equation}
\rho'_{\rm s} = \frac{1}{2\pi} \int_0^{2\pi} d\phi \ \rho_{\rm s}(\phi) ,    
\end{equation}
where here, for clarity, we have written $\rho_{\rm s}(\phi) \coloneqq \hat{D}(\alpha) \rho_{\rm th} \hat{D}(\alpha)^\dagger$ to indicate the explicit dependence of the state of Eq.~\eqref{DisThStt} on the phase $\phi$, where $\alpha = |\alpha| e^{i \phi}$.
Since
\begin{equation}
\langle n|\hat{D}(\alpha)|k \rangle \langle m| \hat{D}(\alpha) |k \rangle^* \propto e^{i\phi (m-n)},
\end{equation}
then
\begin{equation}
\int_0^{2\pi} \langle n|\hat{D}(\alpha)|k \rangle \langle m| \hat{D}(\alpha) |k \rangle^* d\phi \propto \delta_{n,m}.    
\end{equation}
That is, this expression is equal to zero if $n\ne m$. Plugging this observation into Eq.~\eqref{DisThStt2}, we find that the matrix elements of the dephased state $\rho'_{\rm s}$ are given by \cite{lachs1965theoretical}
\begin{multline}
    \langle n| \rho'_{\rm s} |m \rangle =  \frac{\bar n_{\rm th}^n}{(1+\bar n_{\rm th})^{n+1}} \exp\!\left(-\frac{|\alpha|^2}{1+\bar n_{\rm th}}\right)\times \\ L_n\!\left(-\frac{|\alpha|^2}{(1+\bar n_{\rm th})\bar n_{\rm th}}\right) \delta_{n,m}\,.
\end{multline}
Therefore, in the case of an unstable phase for the return signal, the state has no phase information, and a phase-insensitive measurement is optimal. 

We repeat the same analysis for the unstable phase state. The Helstrom bound and the missed-detection error probability bound are plotted with red dashed curves in Figures~\ref{fig:2a} and \ref{fig:2b}, respectively. The simulations are reproduced with one difference; there is no displacement that impacts the photon statistics. The average error probabilities from the simulation are plotted as purple boxes in Figures~\ref{fig:2a} and \ref{fig:2b}. Similar to the known phase case, the optimal symmetric error probability uses the information about the photon statistics of the return signal state. If this use is prevented, the simulation results imply that there is a larger error probability. This is in agreement with the theoretical prediction. 
In a real ranging scenario, the unknown intensity still has a certain value which will dictate an (unknown) error rate based on the values in Figure \ref{fig2}. A bound to the error rate is given by the error of the state with minimal coherent state contribution, which is still considered as a signal (and which is chosen to be $N_s=1$ in here). 

\begin{figure}[t]
    \includegraphics[width=1\columnwidth]{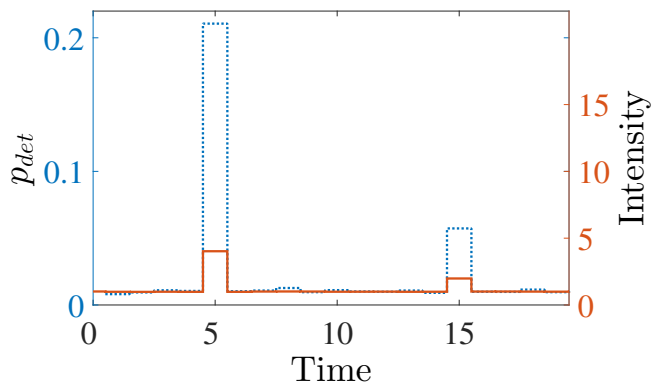}
    \caption{Simulation of a ranging experiment. Intensity (right y-axis) is presented with solid orange line and the optimal detection (left y-axis) with dotted blue line.}
    \label{fig4}
\end{figure}

To demonstrate the theory in the context of time-of-flight ranging, we simulate a time-dependent ranging experiment. The range is divided into 20 discrete slots. A discrete ranging model is based on an experimental consideration in which a finite time resolution of the detection part results in discrete time slots. One target is located at the fifth slot, returning three photons on average, and a second target at the fifteenth slot, returning one photon on average. For all slots, the thermal source contributes one photon on average. The simulation is repeated 10,000 times. Then we apply the decision rule for the asymmetric error setting with false-alarm error threshold, $\varepsilon=0.01$. Figure~\ref{fig4} shows the results for the target detection rate, shown as a dotted blue line. For the slots without a target, the detection rate is $0.0103 \pm 0.001$, agreeing with the set false-alarm error threshold. For the slots with the targets, the detection rate is $0.2106$ and $0.0573$, agreeing with the theoretical missed-detection error probabilities of $0.7833$ and $0.943$, respectively, from Figure~\ref{fig:2b}. The optimal detection is compared to intensity detection. In this example, optimal detection presents 5-7 higher signal relative to the background.

We note here that ancillary Matlab files, which have been used to generate all plots in our paper, are available with the arXiv posting of this paper. 

\section{Discussion} For the setting involving symmetric error probability, the decision rule of \eqref{eq:NP-test} involves two photon distributions. It can be shown that the two distributions intersect only in one point. This means that the distribution with only thermal light is larger up to some photon number \---- the threshold photon number \---- and smaller after it. Thus, no target is presumably detected if a photon number less than the threshold is measured, and the target is presumably detected if a larger photon number is measured. Thus, we have found that the near-optimal detection is photon-number thresholding \cite{cohen2019thresholded} after displacement. For the case of unstable phase, photon number thresholding is optimal. Here, unlike in Ref.~\cite{cohen2019thresholded}, we show how to set the optimal threshold. 

In general, for ranging applications, the setting with asymmetric error probability is more applicable than the symmetric setting. This is due to the lack of a prior probability distribution on the two states and, furthermore, because one is typically willing to tolerate false-detection errors at a fixed rate and then the goal is to minimize the rate of missed-detection errors. The symmetric error setting can be applicable in other situations where the number of return coherent states has a known distribution, for example, in communication \cite{tsujino2011quantum}. 

For the setting with asymmetric error probability, the stochastic decision rule results in a non-deterministic detection strategy, announcing a target with probability $\Omega(n)$, upon measuring $n$ photons.
This detection is very close to threshold detection, since the acceptance probabilities are equal to zero for low photon number and equal to one for high photon number (which also makes the decision deterministic). The only difference is that the value for the intermediate photon number is not an integer. This causes a non-deterministic choice, unlike threshold detection. 
This detection can be thought of as a generalization of a non-integer photon-number threshold detection --- it passes for every detected photon number larger than the threshold photon number, it fails for every photon number lower than the threshold, and passes randomly if the detected photon number equals the threshold. This generalized threshold detection can be realized in the lab by setting the threshold voltage in the middle of the Gaussian packet of the threshold photon number, instead of being in between Gaussian packets of different photon numbers. Thus, when the detected photon number equals the threshold, a target is announced randomly based on the electronic noise, which serves as a random generator.  



This work can be extended in a few directions, for example, exploring if using entanglement gives advantage over coherent state for an unknown return state as it gives for a known state \cite{shapiro2020quantum,zhuang2021quantum}, or alternatively, exploring the limits of simultaneous detection of range and velocity \cite{huang2021quantum} for an unknown return state.

\section{Conclusion}

We have presented the symmetric and asymmetric error probabilities for an unknown return signal, to reveal the optimal yet achievable measurement schemes for ranging applications. We have seen that the quantum limit is unreachable for fixed but unknown phase in a single-shot experiment, since phase information exists but cannot be used. However, for unstable phase, the phase averaging out results in a lower quantum limit. In this case, a single measurement is engineered to saturate the limit, for the asymmetric error setting only. For the symmetric setting, a small gap between the single-measurement errors and the quantum limit is found. This is a result of the unknown intensity. In this point, it is not clear if another measurement can show better performance and this should be resolved in further research. The engineering of this optimal single measurement will lead to the development of quantum-limited rangefinders and LIDARs.   

On the other hand, we have shown that quantum limits of detection might use inaccessible information of the detected state, since typically in quantum information theory, it is often assumed that the receiver has complete information. This results in a limit that is achievable only if the measurement also uses the same information. Since applying such a measurement in practice is impossible, the limit cannot be saturated by any practical measurement. Thus the theory should be modified and achievable bounds should be found \cite{mosonyi2020error,karsa2021energetic}, as has been done here.


\begin{acknowledgements}
L.~C.~and M.~M.~W.~acknowledge support from the Air Force Office of Scientific Research, Grant No.~FA2386-18-1-4010, and thank Kunal Sharma for helpful discussions. 
\end{acknowledgements}

\bibliography{optimal_quantum_sensing}

\providecommand{\noopsort}[1]{}\providecommand{\singleletter}[1]{#1}%
\begin{thebibliography}{32}%
\makeatletter
\providecommand \@ifxundefined [1]{%
 \@ifx{#1\undefined}
}%
\providecommand \@ifnum [1]{%
 \ifnum #1\expandafter \@firstoftwo
 \else \expandafter \@secondoftwo
 \fi
}%
\providecommand \@ifx [1]{%
 \ifx #1\expandafter \@firstoftwo
 \else \expandafter \@secondoftwo
 \fi
}%
\providecommand \natexlab [1]{#1}%
\providecommand \enquote  [1]{``#1''}%
\providecommand \bibnamefont  [1]{#1}%
\providecommand \bibfnamefont [1]{#1}%
\providecommand \citenamefont [1]{#1}%
\providecommand \href@noop [0]{\@secondoftwo}%
\providecommand \href [0]{\begingroup \@sanitize@url \@href}%
\providecommand \@href[1]{\@@startlink{#1}\@@href}%
\providecommand \@@href[1]{\endgroup#1\@@endlink}%
\providecommand \@sanitize@url [0]{\catcode `\\12\catcode `\$12\catcode
  `\&12\catcode `\#12\catcode `\^12\catcode `\_12\catcode `\%12\relax}%
\providecommand \@@startlink[1]{}%
\providecommand \@@endlink[0]{}%
\providecommand \url  [0]{\begingroup\@sanitize@url \@url }%
\providecommand \@url [1]{\endgroup\@href {#1}{\urlprefix }}%
\providecommand \urlprefix  [0]{URL }%
\providecommand \Eprint [0]{\href }%
\providecommand \doibase [0]{https://doi.org/}%
\providecommand \selectlanguage [0]{\@gobble}%
\providecommand \bibinfo  [0]{\@secondoftwo}%
\providecommand \bibfield  [0]{\@secondoftwo}%
\providecommand \translation [1]{[#1]}%
\providecommand \BibitemOpen [0]{}%
\providecommand \bibitemStop [0]{}%
\providecommand \bibitemNoStop [0]{.\EOS\space}%
\providecommand \EOS [0]{\spacefactor3000\relax}%
\providecommand \BibitemShut  [1]{\csname bibitem#1\endcsname}%
\let\auto@bib@innerbib\@empty
\bibitem [{\citenamefont {Hayashi}(2006)}]{H06book}%
  \BibitemOpen
  \bibfield  {author} {\bibinfo {author} {\bibfnamefont {M.}~\bibnamefont
  {Hayashi}},\ }\href@noop {} {\emph {\bibinfo {title} {Quantum Information: An
  Introduction}}}\ (\bibinfo  {publisher} {Springer},\ \bibinfo {year}
  {2006})\BibitemShut {NoStop}%
\bibitem [{\citenamefont {Holevo}(2012)}]{H12}%
  \BibitemOpen
  \bibfield  {author} {\bibinfo {author} {\bibfnamefont {A.~S.}\ \bibnamefont
  {Holevo}},\ }\href@noop {} {\emph {\bibinfo {title} {Quantum Systems,
  Channels, Information}}},\ de Gruyter Studies in Mathematical Physics (Book
  16)\ (\bibinfo  {publisher} {de Gruyter},\ \bibinfo {year} {2012})\ p.\
  \bibinfo {pages} {349}\BibitemShut {NoStop}%
\bibitem [{\citenamefont {Watrous}(2018)}]{Wat16}%
  \BibitemOpen
  \bibfield  {author} {\bibinfo {author} {\bibfnamefont {J.}~\bibnamefont
  {Watrous}},\ }\href@noop {} {\emph {\bibinfo {title} {Theory of Quantum
  Information}}}\ (\bibinfo  {publisher} {Cambridge University Press},\
  \bibinfo {year} {2018})\BibitemShut {NoStop}%
\bibitem [{\citenamefont {Wilde}(2017)}]{W17}%
  \BibitemOpen
  \bibfield  {author} {\bibinfo {author} {\bibfnamefont {M.~M.}\ \bibnamefont
  {Wilde}},\ }\href@noop {} {\emph {\bibinfo {title} {Quantum Information
  Theory}}},\ \bibinfo {edition} {2nd}\ ed.\ (\bibinfo  {publisher} {Cambridge
  University Press},\ \bibinfo {year} {2017})\ \bibinfo {note}
  {arXiv:1106.1445}\BibitemShut {NoStop}%
\bibitem [{\citenamefont {Helstrom}(1969)}]{H69}%
  \BibitemOpen
  \bibfield  {author} {\bibinfo {author} {\bibfnamefont {C.~W.}\ \bibnamefont
  {Helstrom}},\ }\bibfield  {title} {\bibinfo {title} {Quantum detection and
  estimation theory},\ }\href@noop {} {\bibfield  {journal} {\bibinfo
  {journal} {Journal of Statistical Physics}\ }\textbf {\bibinfo {volume}
  {1}},\ \bibinfo {pages} {231} (\bibinfo {year} {1969})}\BibitemShut {NoStop}%
\bibitem [{\citenamefont {Holevo}(1973)}]{japan1973holevo}%
  \BibitemOpen
  \bibfield  {author} {\bibinfo {author} {\bibfnamefont {A.~S.}\ \bibnamefont
  {Holevo}},\ }\bibfield  {title} {\bibinfo {title} {Statistical problems in
  quantum physics},\ }in\ \href@noop {} {\emph {\bibinfo {booktitle} {Second
  Japan-USSR Symposium on Probability Theory}}},\ \bibinfo {series} {Lecture
  Notes in Mathematics}, Vol.\ \bibinfo {volume} {330}\ (\bibinfo  {publisher}
  {Springer Berlin / Heidelberg},\ \bibinfo {year} {1973})\ pp.\ \bibinfo
  {pages} {104--119}\BibitemShut {NoStop}%
\bibitem [{\citenamefont {Helstrom}(1976)}]{Hel76}%
  \BibitemOpen
  \bibfield  {author} {\bibinfo {author} {\bibfnamefont {C.~W.}\ \bibnamefont
  {Helstrom}},\ }\href@noop {} {\emph {\bibinfo {title} {Quantum Detection and
  Estimation Theory}}}\ (\bibinfo  {publisher} {Academic},\ \bibinfo {address}
  {New York},\ \bibinfo {year} {1976})\BibitemShut {NoStop}%
\bibitem [{\citenamefont {Escher}\ \emph {et~al.}(2011)\citenamefont {Escher},
  \citenamefont {de~Matos~Filho},\ and\ \citenamefont
  {Davidovich}}]{escher2011general}%
  \BibitemOpen
  \bibfield  {author} {\bibinfo {author} {\bibfnamefont {B.~M.}\ \bibnamefont
  {Escher}}, \bibinfo {author} {\bibfnamefont {R.~L.}\ \bibnamefont
  {de~Matos~Filho}},\ and\ \bibinfo {author} {\bibfnamefont {L.}~\bibnamefont
  {Davidovich}},\ }\bibfield  {title} {\bibinfo {title} {General framework for
  estimating the ultimate precision limit in noisy quantum-enhanced
  metrology},\ }\href@noop {} {\bibfield  {journal} {\bibinfo  {journal}
  {Nature Physics}\ }\textbf {\bibinfo {volume} {7}},\ \bibinfo {pages} {406}
  (\bibinfo {year} {2011})}\BibitemShut {NoStop}%
\bibitem [{\citenamefont {Degen}\ \emph {et~al.}(2017)\citenamefont {Degen},
  \citenamefont {Reinhard},\ and\ \citenamefont
  {Cappellaro}}]{RevModPhys.89.035002}%
  \BibitemOpen
  \bibfield  {author} {\bibinfo {author} {\bibfnamefont {C.~L.}\ \bibnamefont
  {Degen}}, \bibinfo {author} {\bibfnamefont {F.}~\bibnamefont {Reinhard}},\
  and\ \bibinfo {author} {\bibfnamefont {P.}~\bibnamefont {Cappellaro}},\
  }\bibfield  {title} {\bibinfo {title} {Quantum sensing},\ }\href@noop {}
  {\bibfield  {journal} {\bibinfo  {journal} {Reviews of Modern Physics}\
  }\textbf {\bibinfo {volume} {89}},\ \bibinfo {pages} {035002} (\bibinfo
  {year} {2017})}\BibitemShut {NoStop}%
\bibitem [{\citenamefont {Lloyd}(2008)}]{lloyd2008enhanced}%
  \BibitemOpen
  \bibfield  {author} {\bibinfo {author} {\bibfnamefont {S.}~\bibnamefont
  {Lloyd}},\ }\bibfield  {title} {\bibinfo {title} {Enhanced sensitivity of
  photodetection via quantum illumination},\ }\href@noop {} {\bibfield
  {journal} {\bibinfo  {journal} {Science}\ }\textbf {\bibinfo {volume}
  {321}},\ \bibinfo {pages} {1463} (\bibinfo {year} {2008})}\BibitemShut
  {NoStop}%
\bibitem [{\citenamefont {Shapiro}(2020)}]{shapiro2020quantum}%
  \BibitemOpen
  \bibfield  {author} {\bibinfo {author} {\bibfnamefont {J.~H.}\ \bibnamefont
  {Shapiro}},\ }\bibfield  {title} {\bibinfo {title} {The quantum illumination
  story},\ }\href@noop {} {\bibfield  {journal} {\bibinfo  {journal} {IEEE
  Aerospace and Electronic Systems Magazine}\ }\textbf {\bibinfo {volume}
  {35}},\ \bibinfo {pages} {8} (\bibinfo {year} {2020})}\BibitemShut {NoStop}%
\bibitem [{\citenamefont {Zhuang}\ \emph {et~al.}(2017)\citenamefont {Zhuang},
  \citenamefont {Zhang},\ and\ \citenamefont
  {Shapiro}}]{zhuang2017entanglement}%
  \BibitemOpen
  \bibfield  {author} {\bibinfo {author} {\bibfnamefont {Q.}~\bibnamefont
  {Zhuang}}, \bibinfo {author} {\bibfnamefont {Z.}~\bibnamefont {Zhang}},\ and\
  \bibinfo {author} {\bibfnamefont {J.~H.}\ \bibnamefont {Shapiro}},\
  }\bibfield  {title} {\bibinfo {title} {Entanglement-enhanced
  {Neyman--Pearson} target detection using quantum illumination},\ }\href@noop
  {} {\bibfield  {journal} {\bibinfo  {journal} {Journal of the Optical Society
  of America B}\ }\textbf {\bibinfo {volume} {34}},\ \bibinfo {pages} {1567}
  (\bibinfo {year} {2017})}\BibitemShut {NoStop}%
\bibitem [{\citenamefont {Gerry}\ and\ \citenamefont {Knight}(2004)}]{GK04}%
  \BibitemOpen
  \bibfield  {author} {\bibinfo {author} {\bibfnamefont {C.}~\bibnamefont
  {Gerry}}\ and\ \bibinfo {author} {\bibfnamefont {P.}~\bibnamefont {Knight}},\
  }\href@noop {} {\emph {\bibinfo {title} {Introductory Quantum Optics}}}\
  (\bibinfo  {publisher} {Cambridge University Press},\ \bibinfo {year}
  {2004})\ p.\ \bibinfo {pages} {332}\BibitemShut {NoStop}%
\bibitem [{\citenamefont {Ivan}\ \emph {et~al.}(2011)\citenamefont {Ivan},
  \citenamefont {Sabapathy},\ and\ \citenamefont {Simon}}]{ivan2011operator}%
  \BibitemOpen
  \bibfield  {author} {\bibinfo {author} {\bibfnamefont {J.~S.}\ \bibnamefont
  {Ivan}}, \bibinfo {author} {\bibfnamefont {K.~K.}\ \bibnamefont
  {Sabapathy}},\ and\ \bibinfo {author} {\bibfnamefont {R.}~\bibnamefont
  {Simon}},\ }\bibfield  {title} {\bibinfo {title} {Operator-sum representation
  for bosonic {Gaussian} channels},\ }\href@noop {} {\bibfield  {journal}
  {\bibinfo  {journal} {Physical Review A}\ }\textbf {\bibinfo {volume} {84}},\
  \bibinfo {pages} {042311} (\bibinfo {year} {2011})}\BibitemShut {NoStop}%
\bibitem [{\citenamefont {Helstrom}(1967)}]{Hel67}%
  \BibitemOpen
  \bibfield  {author} {\bibinfo {author} {\bibfnamefont {C.~W.}\ \bibnamefont
  {Helstrom}},\ }\bibfield  {title} {\bibinfo {title} {Detection theory and
  quantum mechanics},\ }\href@noop {} {\bibfield  {journal} {\bibinfo
  {journal} {Information and Control}\ }\textbf {\bibinfo {volume} {10}},\
  \bibinfo {pages} {254} (\bibinfo {year} {1967})}\BibitemShut {NoStop}%
\bibitem [{\citenamefont {Kennedy}(1973)}]{kennedy1973near}%
  \BibitemOpen
  \bibfield  {author} {\bibinfo {author} {\bibfnamefont {R.~S.}\ \bibnamefont
  {Kennedy}},\ }\bibfield  {title} {\bibinfo {title} {A near-optimum receiver
  for the binary coherent state quantum channel},\ }\href@noop {} {\bibfield
  {journal} {\bibinfo  {journal} {Quarterly Progress Report}\ }\textbf
  {\bibinfo {volume} {108}},\ \bibinfo {pages} {219} (\bibinfo {year}
  {1973})}\BibitemShut {NoStop}%
\bibitem [{\citenamefont {Takeoka}\ and\ \citenamefont
  {Sasaki}(2008)}]{takeoka2008discrimination}%
  \BibitemOpen
  \bibfield  {author} {\bibinfo {author} {\bibfnamefont {M.}~\bibnamefont
  {Takeoka}}\ and\ \bibinfo {author} {\bibfnamefont {M.}~\bibnamefont
  {Sasaki}},\ }\bibfield  {title} {\bibinfo {title} {Discrimination of the
  binary coherent signal: Gaussian-operation limit and simple non-gaussian
  near-optimal receivers},\ }\href@noop {} {\bibfield  {journal} {\bibinfo
  {journal} {Physical Review A}\ }\textbf {\bibinfo {volume} {78}},\ \bibinfo
  {pages} {022320} (\bibinfo {year} {2008})}\BibitemShut {NoStop}%
\bibitem [{\citenamefont {Yoshitani}(1970)}]{Yoshitani1970}%
  \BibitemOpen
  \bibfield  {author} {\bibinfo {author} {\bibfnamefont {R.}~\bibnamefont
  {Yoshitani}},\ }\bibfield  {title} {\bibinfo {title} {On the detectabillity
  limit of coherent optical signals in thermal radiation},\ }\href@noop {}
  {\bibfield  {journal} {\bibinfo  {journal} {Journal of Statistical Physics}\
  }\textbf {\bibinfo {volume} {2}},\ \bibinfo {pages} {347} (\bibinfo {year}
  {1970})}\BibitemShut {NoStop}%
\bibitem [{\citenamefont {Habif}\ \emph {et~al.}(2021)\citenamefont {Habif},
  \citenamefont {Jagannathan}, \citenamefont {Gartenstein}, \citenamefont
  {Amory},\ and\ \citenamefont {Guha}}]{habif2021quantum}%
  \BibitemOpen
  \bibfield  {author} {\bibinfo {author} {\bibfnamefont {J.~L.}\ \bibnamefont
  {Habif}}, \bibinfo {author} {\bibfnamefont {A.}~\bibnamefont {Jagannathan}},
  \bibinfo {author} {\bibfnamefont {S.}~\bibnamefont {Gartenstein}}, \bibinfo
  {author} {\bibfnamefont {P.}~\bibnamefont {Amory}},\ and\ \bibinfo {author}
  {\bibfnamefont {S.}~\bibnamefont {Guha}},\ }\bibfield  {title} {\bibinfo
  {title} {Quantum-limited discrimination of laser light and thermal light},\
  }\href@noop {} {\bibfield  {journal} {\bibinfo  {journal} {Optics Express}\
  }\textbf {\bibinfo {volume} {29}},\ \bibinfo {pages} {7418} (\bibinfo {year}
  {2021})}\BibitemShut {NoStop}%
\bibitem [{\citenamefont {You}\ \emph {et~al.}(2020)\citenamefont {You},
  \citenamefont {Quiroz-Ju{\'a}rez}, \citenamefont {Lambert}, \citenamefont
  {Bhusal}, \citenamefont {Dong}, \citenamefont {Perez-Leija}, \citenamefont
  {Javaid}, \citenamefont {Le{\'o}n-Montiel},\ and\ \citenamefont
  {Maga{\~n}a-Loaiza}}]{you2020identification}%
  \BibitemOpen
  \bibfield  {author} {\bibinfo {author} {\bibfnamefont {C.}~\bibnamefont
  {You}}, \bibinfo {author} {\bibfnamefont {M.~A.}\ \bibnamefont
  {Quiroz-Ju{\'a}rez}}, \bibinfo {author} {\bibfnamefont {A.}~\bibnamefont
  {Lambert}}, \bibinfo {author} {\bibfnamefont {N.}~\bibnamefont {Bhusal}},
  \bibinfo {author} {\bibfnamefont {C.}~\bibnamefont {Dong}}, \bibinfo {author}
  {\bibfnamefont {A.}~\bibnamefont {Perez-Leija}}, \bibinfo {author}
  {\bibfnamefont {A.}~\bibnamefont {Javaid}}, \bibinfo {author} {\bibfnamefont
  {R.~D.~J.}\ \bibnamefont {Le{\'o}n-Montiel}},\ and\ \bibinfo {author}
  {\bibfnamefont {O.~S.}\ \bibnamefont {Maga{\~n}a-Loaiza}},\ }\bibfield
  {title} {\bibinfo {title} {Identification of light sources using machine
  learning},\ }\href@noop {} {\bibfield  {journal} {\bibinfo  {journal}
  {Applied Physics Reviews}\ }\textbf {\bibinfo {volume} {7}},\ \bibinfo
  {pages} {021404} (\bibinfo {year} {2020})}\BibitemShut {NoStop}%
\bibitem [{\citenamefont {Sher}\ \emph {et~al.}(2018)\citenamefont {Sher},
  \citenamefont {Cohen}, \citenamefont {Istrati},\ and\ \citenamefont
  {Eisenberg}}]{sher2018low}%
  \BibitemOpen
  \bibfield  {author} {\bibinfo {author} {\bibfnamefont {Y.}~\bibnamefont
  {Sher}}, \bibinfo {author} {\bibfnamefont {L.}~\bibnamefont {Cohen}},
  \bibinfo {author} {\bibfnamefont {D.}~\bibnamefont {Istrati}},\ and\ \bibinfo
  {author} {\bibfnamefont {H.~S.}\ \bibnamefont {Eisenberg}},\ }\bibfield
  {title} {\bibinfo {title} {Low intensity {LiDAR} using compressed sensing and
  a photon number resolving detector},\ }in\ \href@noop {} {\emph {\bibinfo
  {booktitle} {Emerging Digital Micromirror Device Based Systems and
  Applications X}}},\ Vol.\ \bibinfo {volume} {10546}\ (\bibinfo {organization}
  {International Society for Optics and Photonics},\ \bibinfo {year} {2018})\
  p.\ \bibinfo {pages} {105460J}\BibitemShut {NoStop}%
\bibitem [{\citenamefont {Trees}(2001)}]{vT01book}%
  \BibitemOpen
  \bibfield  {author} {\bibinfo {author} {\bibfnamefont {H.~L.~V.}\
  \bibnamefont {Trees}},\ }\href@noop {} {\emph {\bibinfo {title} {Detection,
  Estimation, and Modulation Theory, Part I: Detection, Estimation, and Linear
  Modulation Theory}}}\ (\bibinfo  {publisher} {John Wiley \& Sons, Ltd.},\
  \bibinfo {year} {2001})\BibitemShut {NoStop}%
\bibitem [{\citenamefont {Wilde}\ \emph {et~al.}(2017)\citenamefont {Wilde},
  \citenamefont {Tomamichel}, \citenamefont {Lloyd},\ and\ \citenamefont
  {Berta}}]{wilde2017gaussian}%
  \BibitemOpen
  \bibfield  {author} {\bibinfo {author} {\bibfnamefont {M.~M.}\ \bibnamefont
  {Wilde}}, \bibinfo {author} {\bibfnamefont {M.}~\bibnamefont {Tomamichel}},
  \bibinfo {author} {\bibfnamefont {S.}~\bibnamefont {Lloyd}},\ and\ \bibinfo
  {author} {\bibfnamefont {M.}~\bibnamefont {Berta}},\ }\bibfield  {title}
  {\bibinfo {title} {Gaussian hypothesis testing and quantum illumination},\
  }\href@noop {} {\bibfield  {journal} {\bibinfo  {journal} {Physical Review
  Letters}\ }\textbf {\bibinfo {volume} {119}},\ \bibinfo {pages} {120501}
  (\bibinfo {year} {2017})}\BibitemShut {NoStop}%
\bibitem [{\citenamefont {Dupuis}\ \emph {et~al.}(2013)\citenamefont {Dupuis},
  \citenamefont {Kraemer}, \citenamefont {Faist}, \citenamefont {Renes},\ and\
  \citenamefont {Renner}}]{DKFRR13}%
  \BibitemOpen
  \bibfield  {author} {\bibinfo {author} {\bibfnamefont {F.}~\bibnamefont
  {Dupuis}}, \bibinfo {author} {\bibfnamefont {L.}~\bibnamefont {Kraemer}},
  \bibinfo {author} {\bibfnamefont {P.}~\bibnamefont {Faist}}, \bibinfo
  {author} {\bibfnamefont {J.~M.}\ \bibnamefont {Renes}},\ and\ \bibinfo
  {author} {\bibfnamefont {R.}~\bibnamefont {Renner}},\ }\bibfield  {title}
  {\bibinfo {title} {Generalized entropies},\ }\href@noop {} {\bibfield
  {journal} {\bibinfo  {journal} {XVIIth International Congress on Mathematical
  Physics}\ ,\ \bibinfo {pages} {134}} (\bibinfo {year} {2013})},\ \bibinfo
  {note} {arXiv:1211.3141}\BibitemShut {NoStop}%
\bibitem [{\citenamefont {Lachs}(1965)}]{lachs1965theoretical}%
  \BibitemOpen
  \bibfield  {author} {\bibinfo {author} {\bibfnamefont {G.}~\bibnamefont
  {Lachs}},\ }\bibfield  {title} {\bibinfo {title} {Theoretical aspects of
  mixtures of thermal and coherent radiation},\ }\href@noop {} {\bibfield
  {journal} {\bibinfo  {journal} {Physical Review}\ }\textbf {\bibinfo {volume}
  {138}},\ \bibinfo {pages} {B1012} (\bibinfo {year} {1965})}\BibitemShut
  {NoStop}%
\bibitem [{\citenamefont {Cohen}\ \emph {et~al.}(2019)\citenamefont {Cohen},
  \citenamefont {Matekole}, \citenamefont {Sher}, \citenamefont {Istrati},
  \citenamefont {Eisenberg},\ and\ \citenamefont
  {Dowling}}]{cohen2019thresholded}%
  \BibitemOpen
  \bibfield  {author} {\bibinfo {author} {\bibfnamefont {L.}~\bibnamefont
  {Cohen}}, \bibinfo {author} {\bibfnamefont {E.~S.}\ \bibnamefont {Matekole}},
  \bibinfo {author} {\bibfnamefont {Y.}~\bibnamefont {Sher}}, \bibinfo {author}
  {\bibfnamefont {D.}~\bibnamefont {Istrati}}, \bibinfo {author} {\bibfnamefont
  {H.~S.}\ \bibnamefont {Eisenberg}},\ and\ \bibinfo {author} {\bibfnamefont
  {J.~P.}\ \bibnamefont {Dowling}},\ }\bibfield  {title} {\bibinfo {title}
  {Thresholded quantum {LIDAR}: exploiting photon-number-resolving detection},\
  }\href@noop {} {\bibfield  {journal} {\bibinfo  {journal} {Physical Review
  Letters}\ }\textbf {\bibinfo {volume} {123}},\ \bibinfo {pages} {203601}
  (\bibinfo {year} {2019})}\BibitemShut {NoStop}%
\bibitem [{\citenamefont {Tsujino}\ \emph {et~al.}(2011)\citenamefont
  {Tsujino}, \citenamefont {Fukuda}, \citenamefont {Fujii}, \citenamefont
  {Inoue}, \citenamefont {Fujiwara}, \citenamefont {Takeoka},\ and\
  \citenamefont {Sasaki}}]{tsujino2011quantum}%
  \BibitemOpen
  \bibfield  {author} {\bibinfo {author} {\bibfnamefont {K.}~\bibnamefont
  {Tsujino}}, \bibinfo {author} {\bibfnamefont {D.}~\bibnamefont {Fukuda}},
  \bibinfo {author} {\bibfnamefont {G.}~\bibnamefont {Fujii}}, \bibinfo
  {author} {\bibfnamefont {S.}~\bibnamefont {Inoue}}, \bibinfo {author}
  {\bibfnamefont {M.}~\bibnamefont {Fujiwara}}, \bibinfo {author}
  {\bibfnamefont {M.}~\bibnamefont {Takeoka}},\ and\ \bibinfo {author}
  {\bibfnamefont {M.}~\bibnamefont {Sasaki}},\ }\bibfield  {title} {\bibinfo
  {title} {Quantum receiver beyond the standard quantum limit of coherent
  optical communication},\ }\href@noop {} {\bibfield  {journal} {\bibinfo
  {journal} {Physical Review Letters}\ }\textbf {\bibinfo {volume} {106}},\
  \bibinfo {pages} {250503} (\bibinfo {year} {2011})}\BibitemShut {NoStop}%
\bibitem [{\citenamefont {Zhuang}(2021)}]{zhuang2021quantum}%
  \BibitemOpen
  \bibfield  {author} {\bibinfo {author} {\bibfnamefont {Q.}~\bibnamefont
  {Zhuang}},\ }\bibfield  {title} {\bibinfo {title} {Quantum ranging with
  gaussian entanglement},\ }\href@noop {} {\bibfield  {journal} {\bibinfo
  {journal} {Physical Review Letters}\ }\textbf {\bibinfo {volume} {126}},\
  \bibinfo {pages} {240501} (\bibinfo {year} {2021})}\BibitemShut {NoStop}%
\bibitem [{\citenamefont {Huang}\ \emph {et~al.}(2021)\citenamefont {Huang},
  \citenamefont {Lupo},\ and\ \citenamefont {Kok}}]{huang2021quantum}%
  \BibitemOpen
  \bibfield  {author} {\bibinfo {author} {\bibfnamefont {Z.}~\bibnamefont
  {Huang}}, \bibinfo {author} {\bibfnamefont {C.}~\bibnamefont {Lupo}},\ and\
  \bibinfo {author} {\bibfnamefont {P.}~\bibnamefont {Kok}},\ }\bibfield
  {title} {\bibinfo {title} {Quantum-limited estimation of range and
  velocity},\ }\href@noop {} {\bibfield  {journal} {\bibinfo  {journal} {PRX
  Quantum}\ }\textbf {\bibinfo {volume} {2}},\ \bibinfo {pages} {030303}
  (\bibinfo {year} {2021})}\BibitemShut {NoStop}%
\bibitem [{\citenamefont {Mosonyi}\ \emph {et~al.}(2020)\citenamefont
  {Mosonyi}, \citenamefont {Szil{\'a}gyi},\ and\ \citenamefont
  {Weiner}}]{mosonyi2020error}%
  \BibitemOpen
  \bibfield  {author} {\bibinfo {author} {\bibfnamefont {M.}~\bibnamefont
  {Mosonyi}}, \bibinfo {author} {\bibfnamefont {Z.}~\bibnamefont
  {Szil{\'a}gyi}},\ and\ \bibinfo {author} {\bibfnamefont {M.}~\bibnamefont
  {Weiner}},\ }\bibfield  {title} {\bibinfo {title} {On the error exponents of
  binary quantum state discrimination with composite hypotheses},\ }\href
  {https://arxiv.org/abs/2011.04645} {\bibfield  {journal} {\bibinfo  {journal}
  {arXiv preprint arXiv:2011.04645}\ } (\bibinfo {year} {2020})}\BibitemShut
  {NoStop}%
\bibitem [{\citenamefont {Karsa}\ and\ \citenamefont
  {Pirandola}(2021)}]{karsa2021energetic}%
  \BibitemOpen
  \bibfield  {author} {\bibinfo {author} {\bibfnamefont {A.}~\bibnamefont
  {Karsa}}\ and\ \bibinfo {author} {\bibfnamefont {S.}~\bibnamefont
  {Pirandola}},\ }\bibfield  {title} {\bibinfo {title} {Energetic
  considerations in quantum target ranging},\ }in\ \href@noop {} {\emph
  {\bibinfo {booktitle} {2021 IEEE Radar Conference (RadarConf21)}}}\ (\bibinfo
  {organization} {IEEE},\ \bibinfo {year} {2021})\ pp.\ \bibinfo {pages}
  {1--4}\BibitemShut {NoStop}%
\bibitem [{\citenamefont {Cohen}\ \emph {et~al.}(2018)\citenamefont {Cohen},
  \citenamefont {Pilnyak}, \citenamefont {Istrati}, \citenamefont {Studer},
  \citenamefont {Dowling},\ and\ \citenamefont
  {Eisenberg}}]{cohen2018absolute}%
  \BibitemOpen
  \bibfield  {author} {\bibinfo {author} {\bibfnamefont {L.}~\bibnamefont
  {Cohen}}, \bibinfo {author} {\bibfnamefont {Y.}~\bibnamefont {Pilnyak}},
  \bibinfo {author} {\bibfnamefont {D.}~\bibnamefont {Istrati}}, \bibinfo
  {author} {\bibfnamefont {N.~M.}\ \bibnamefont {Studer}}, \bibinfo {author}
  {\bibfnamefont {J.~P.}\ \bibnamefont {Dowling}},\ and\ \bibinfo {author}
  {\bibfnamefont {H.~S.}\ \bibnamefont {Eisenberg}},\ }\bibfield  {title}
  {\bibinfo {title} {Absolute calibration of single-photon and multiplexed
  photon-number-resolving detectors},\ }\href@noop {} {\bibfield  {journal}
  {\bibinfo  {journal} {Physical Review A}\ }\textbf {\bibinfo {volume} {98}},\
  \bibinfo {pages} {013811} (\bibinfo {year} {2018})}\BibitemShut {NoStop}%
\end{thebibliography}%

\appendix

\section{Phase sensitive measurement for unknown phase}
\label{appendix:a}
\begin{figure}[b]
    \includegraphics[width=0.9\columnwidth]{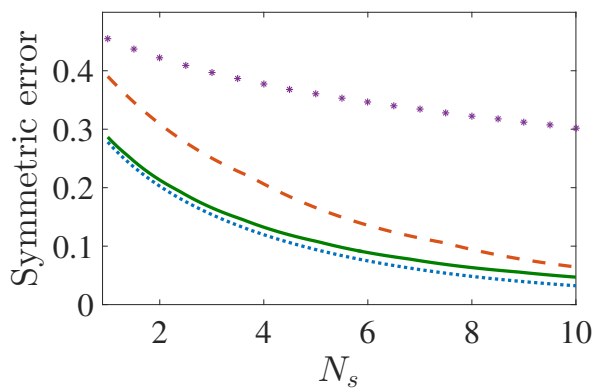}
    \caption{Symmetric error probability. The blue dotted curve is the Helstrom bound. The limit of photon-number detection with and without optimal displacement is presented with a green solid curve and a red dashed curve, respectively. The average error for the optimal displacement when integrating over the signal phase is shown with purple stars.}
    \label{figs2}
\end{figure}

For the known phase case, we found that applying an optimal displacement followed by photon-number detection leads to an error probability that is near the quantum limit (green solid and blue dotted curves in Figure~~\textcolor{blue}{\ref{figs2}}). This detection strategy is phase sensitive and works well when the phase is fixed and known, but less well for the unstable phase case. The error of this strategy while averaging over all signal phases is shown in Figure~\textcolor{blue}{\ref{figs2}} by purple stars. It is clear that the average error is much worse than when applying a phase-insensitive measurement, presented by a red dashed curve.

\section{Decision rule for the asymmetric error setting}
\label{appendix:b}

\begin{figure}[bt]
    \includegraphics[width=0.9\columnwidth]{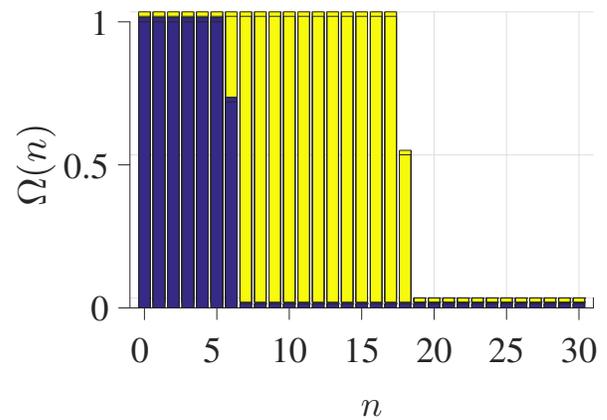}
    \caption{Acceptance probabilities as a function of measured photon number. The rear yellow bars are the acceptance probabilities for the simulated data of Figure~\textcolor{blue}{1a} in the main text, and the front blue bars are the acceptance probabilities for the simulated data of Figure~\textcolor{blue}{2a} in the main text. See main text for more details.}
    \label{figs3}
\end{figure}

For the setting with asymmetric error probability, the stochastic decision rule is found in Eq.~\eqref{unbalErr} of the main text. The acceptance probabilities, $\Omega(n)$, upon measuring $n$ photons are displayed in Figure~\textcolor{blue}{\ref{figs3}}.

\section{Imperfect detector}
\label{appendix:c}

A practical experimental setup suffers from imperfections. The protocol presented in the main text can be easily modified to fit modeled imperfections. Usually, optical loss is the most significant effect. However, here it does not affect our results, given that a displaced thermal state remains such after loss. All the above discussion is then conducted on the measured state, i.e., after loss. Other imperfections will change the measured photon statistics \cite{cohen2018absolute}. Since this change is modeled and known, the rest of the detection protocol stays the same but then is conducted with the modified statistics.

\section{Photon resolution effect}
\label{appendix:d}

\begin{figure}[tb]
    \includegraphics[width=0.9\columnwidth]{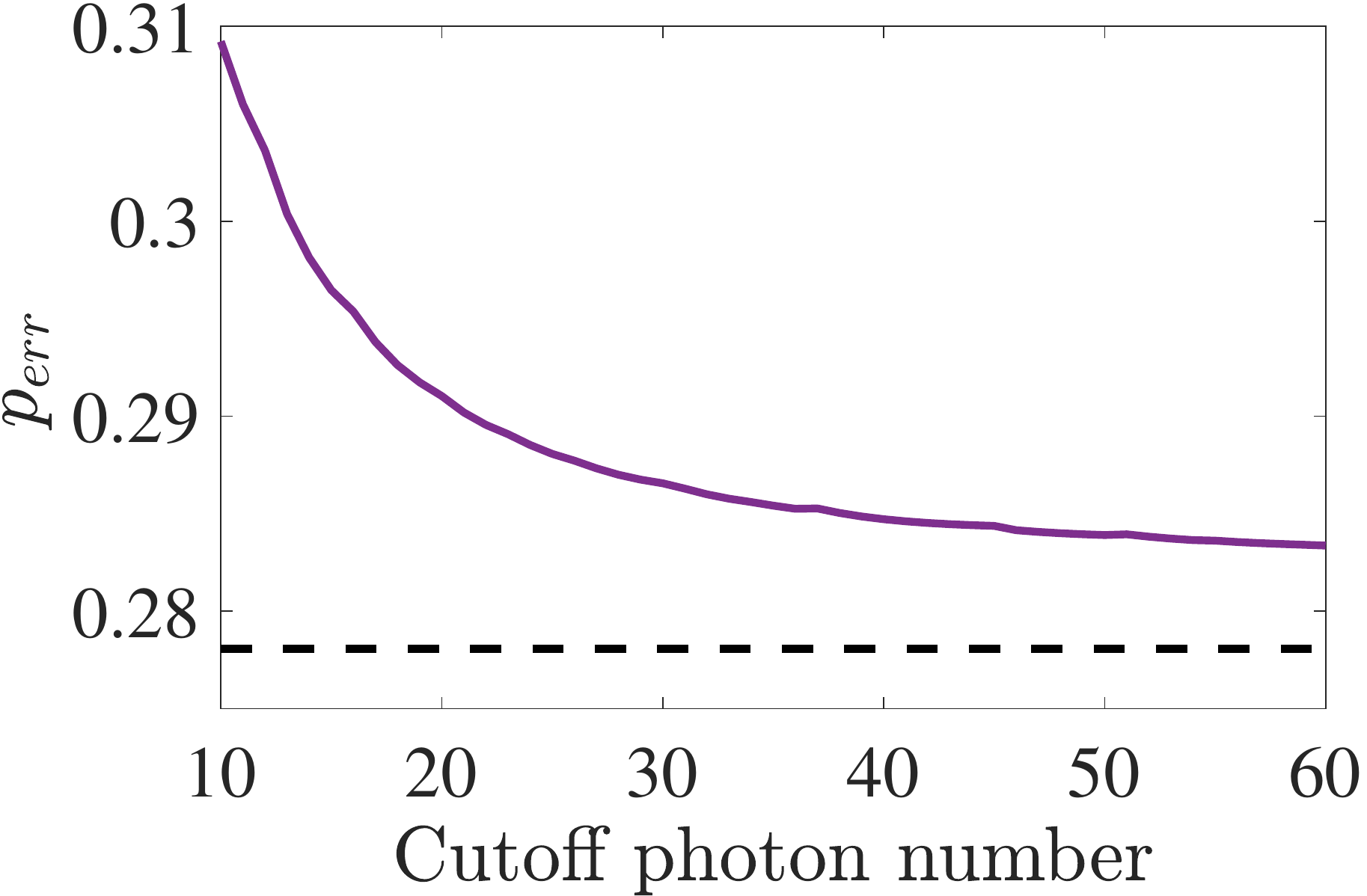}
    \caption{Symmetric error probability as a function of cutoff photon number, for $N_s = \bar n_{\rm th} = 1$. The black dashed line is the Helstrom bound. Using higher photon-number resolution enables to intensify the displacement and to get closer to the Helstrom bound.}
    \label{figs1}
\end{figure}

The optimal displacement is highly sensitive to the chosen cutoff photon number. Increasing the cutoff number increases the optimal displacement parameter and decreases the error probability. The error probability as a function of the cutoff number is depicted in Figure~\textcolor{blue}{\ref{figs1}}. We speculate that in the limit of infinite displacement (and infinite photon-number resolution), this measurement scheme saturates the quantum bound. In the main text, we have employed a cutoff of 30 photons.


\end{document}